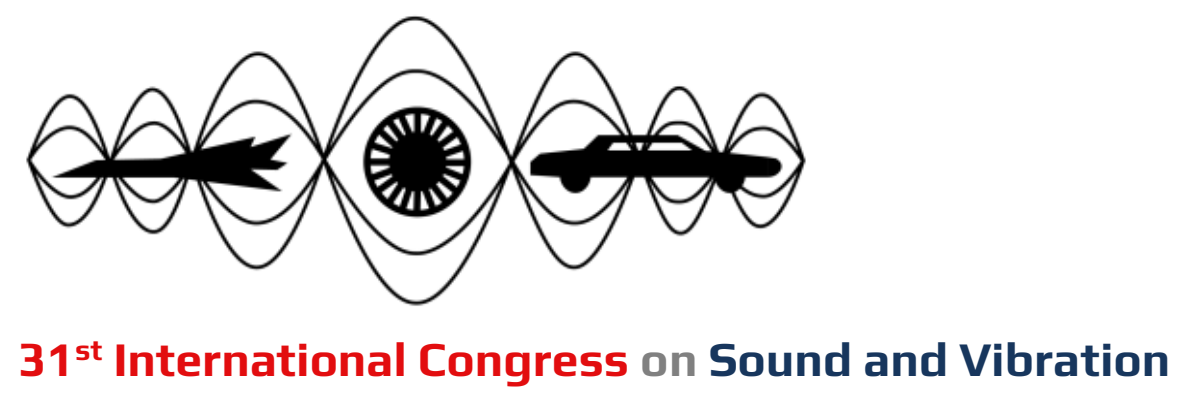

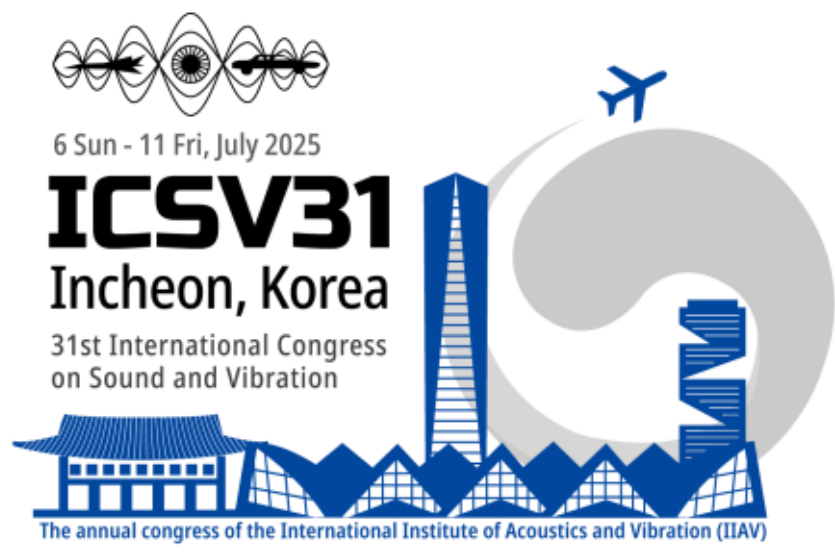

# DEEP LEARNING-BASED ACTIVE TRIM PANELS FOR ENHANCED AIRCRAFT INTERIOR NOISE CONTROL

Boxiang Wang, Zhengding Luo, Junwei Ji, Xiaoyi Shen, Dongyuan Shi and Woon-Seng Gan
*School of Electrical and Electronic Engineering, Nanyang Technological University, Singapore*
*email: boxiang001@e.ntu.edu.sg*

Malte Misol
*German Aerospace Center, Institute of Lightweight Systems, Germany*

Active noise control (ANC) trim panels offer an effective solution to suppress multi-tonal noise in aircraft. The selective fixed-filter ANC (SFANC) method, characterized by low computational complexity, high robustness and rapid response, is suitable to handle multi-tonal engine noise that varies in frequency due to changes in the rotational speed of the engine shaft. However, real-world conditions introduce variations in lining temperature, altering acoustic and structural paths and degrading noise reduction performance. To address this challenge, a temperature-perceptive SFANC (TP-SFANC) approach is proposed that employs a lightweight one-dimensional convolutional neural network (1D CNN) trained using a multi-task learning strategy. By processing both reference and error signals, the 1D CNN learns frequency and temperature characteristics to dynamically select the optimal control filter. Numerical simulations demonstrate the effectiveness of the proposed method in attenuating multi-tonal noise across varying frequencies and lining temperatures.

*Keywords: active noise control, convolutional neural network, multi-task learning, aircraft*

---

## 1. Introduction

The control of aircraft interior noise has been a significant research focus for decades. Active noise control (ANC), particularly effective at low frequencies, has emerged as a promising alternative to traditional passive methods [1, 2]. By generating an anti-noise signal with same amplitude but opposite phase, ANC can substantially reduce sound pressure. Among various algorithms, the filtered-x least mean square (FxLMS) is the most widely used due to its adaptability to changing acoustic environments [3]. Early studies demonstrated the feasibility of ANC in aircraft cabins, with Zalas and Tichy achieving 7–15 dB noise reduction in turboprop aircraft [4], and Elliott et al. [5] reporting over 10 dB reduction in B.Ae. 748 aircraft. To overcome the limitations of loudspeakers as secondary sources, researchers have explored using actuators in active trim panels to directly excite interior structures [6–8]. These panels are lightweight, energy-efficient, and easier to implement than active fuselage structures, offering a promising solution for aircraft interior noise control [8].

Multi-tonal engine noise varies in frequency due to changes in the rotational speed of the engine shaft [9]. The selective fixed-filter ANC (SFANC) method is capable of selecting a suitable control filter for primary noise with varying frequency characteristics. It improves upon the FxLMS algorithm by offering faster response, greater stability, and lower computational complexity [10-13], making it well-suited for controlling multi-tonal engine noise. However, in addition to noise characteristics, temperature variations in the trim panels can significantly alter acoustic and structural paths, thereby degrading the performance of the SFANC technique. Therefore, it is essential to develop a more advanced technique that accounts for both noise characteristics and temperature-induced path variations for aircraft interior noise control.

To address this, this paper proposes a temperature-perceptive SFANC (TP-SFANC) technique using a lightweight (convolutional neural network) CNN that selects the optimal control filter based on reference and error signals under varying noise and temperature conditions. The remainder of this paper is organized as follows: Section 2 describes the experimental setup, Section 3 details the TP-SFANC method, Section 4 presents simulation results, and Section 5 concludes with future research directions.

## 2. Experimental setup

### 2.1 Double panel system with CFRP fuselage and sidewall panel

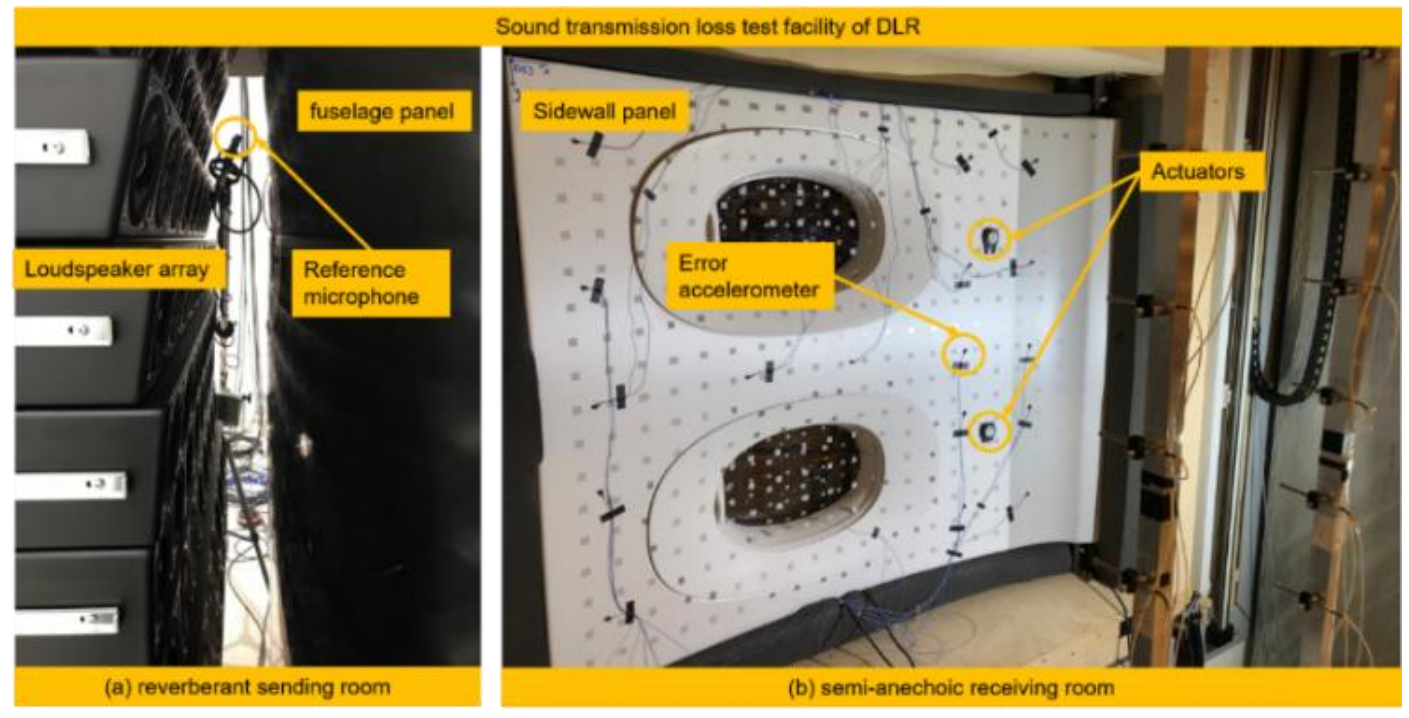

Figure 1: Experimental setup in sound transmission loss test facility with (a) sending room representing the aircraft exterior and (b) receiving room representing the aircraft interior.

Figure 1 shows the experimental setup in the sound transmission loss test facility with a sidewall panel coupled to a carbon fibre reinforced plastics (CFRP) fuselage structure. The setup is representative for an Airbus A350 aircraft. The loudspeaker array (LSA) located in the sending room is placed in front of the fuselage structure leaving a gap of approximately 0.15 m. 72 loudspeakers are driven in parallel to achieve a high sound pressure level (SPL) in front of the fuselage structure. A microphone is placed in the gap between the LSA and the CFRP panel to capture the sound pressure and to provide a reference signal for the ANC system. The sidewall panel is equipped with two inertia actuators of the type Visaton EX45S and with accelerometers measuring the normal surface vibration at different locations. For the current investigation the one accelerometer indicated in Fig. 1(b) is used as the error sensor of the ANC system. This leads to a 1×2×1 configuration with 1 reference microphone, 2 actuators and 1 error sensor.

### 2.2 Counter rotating open rotor engine noise synthesis

A multi-tonal noise signal is generated containing the first five harmonics of a generic counter rotating open rotor (CROR) engine. The CROR engine is considered representative for other energy efficient engines such as turboprops, propfans and unducted fan engines. More information on the CROR noise source and the generation of control voltages for the LSA is provided in [9]. In order to account for broadband noise components coming from the turbulent boundary layer (TBL), the power spectral density of a TBL in a low-speed configuration is derived from a Goody model [14]. The generated noise signal shown in Fig. 2 contains five harmonics: $f_1 = 119.65$ Hz, $f_2 = 1.25 f_1$, $f_3 = f_1 + f_2$, $f_4 = 2 f_1 + f_2$ and $f_5 = f_1 + 2f_2$. The overlaid TBL noise floor has a PSD of 70 dB/Hz at 200 Hz. This combined noise signal serves as a reference signal for the ANC system entering the primary path. Similarly, the multi-narrowband noise can be constructed by treating $f_1$ as a noise signal with a certain bandwidth.

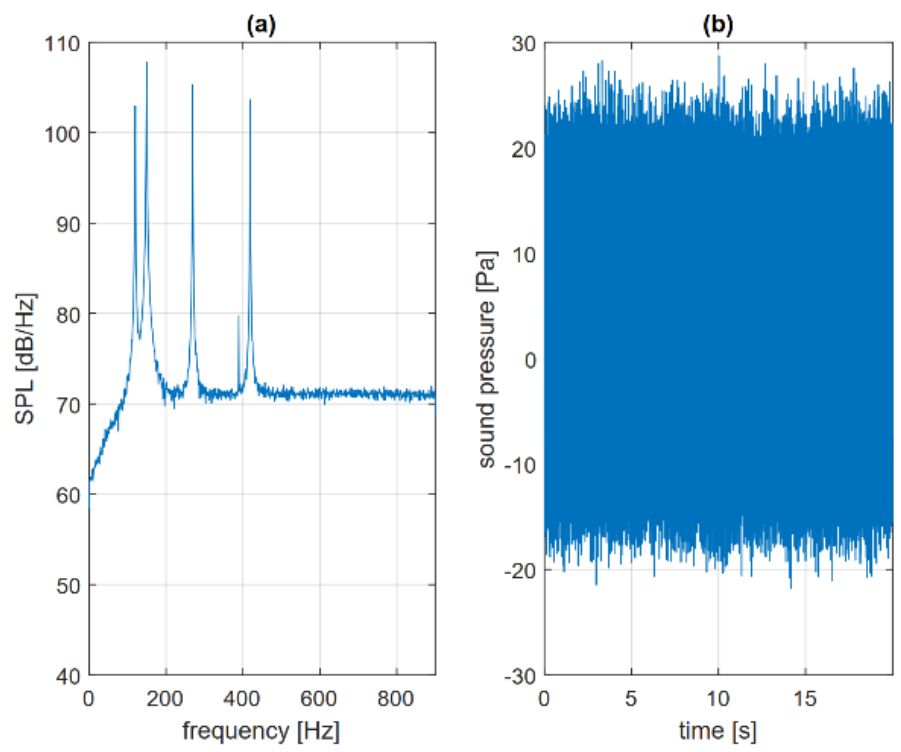

Figure 2: Combined CROR and TBL noise signal in (a) frequency and (b) time domain.

## 2.3 Temperature-induced path variations synthesis

It was shown in [15] that the structural dynamics of the sidewall panel undergo variations depending on the temperature. The operating temperature range for linings from EUROCAE ED-14G is -15°C to 55°C. Temperatures from 22°C to 50°C could be reached in the laboratory by means of an infrared heating device. The changes in the frequency response functions can exceed 10 dB in magnitude and 0.9 rad in phase angle. The changes are mainly attributed to the temperature dependency of the bending stiffness leading to a shift in the eigenfrequencies of the sidewall panel structure. The eigenfrequencies continuously drop for increasing temperatures because of the structural softening. This effect is exploited to generate synthetic models of the physical plant for arbitrary temperatures by correlating the eigenfrequency with the temperature. A state space model for a certain temperature is derived from a single baseline measurement by scaling the original sampling frequency down (for increasing temperatures) or up (for decreasing temperatures). The scaling factor is -0.280366 % per Kelvin.

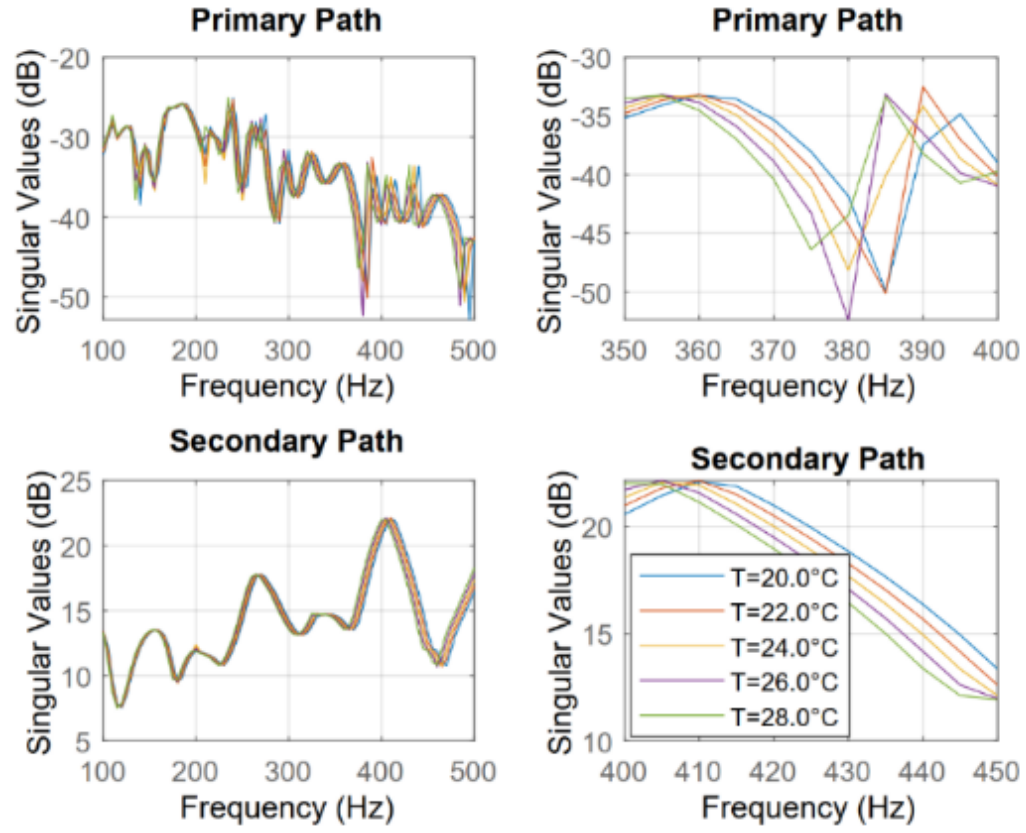

Figure 3: Singular value of primary and secondary paths for the temperature range.

The temperature dependence of the singular values of the primary and secondary paths is shown in Fig. 3. The effect of temperature on the eigenfrequencies is clearly visible. Since every plant model is obtained from an independent subspace identification run, some additional deviations are visible in the singular values that are not attributed to the temperature change but to the approximative nature of the system identification itself.

# 3. Proposed temperature-perceptive SFANC technique

The block diagram of the proposed TP-SFANC technique is presented in Fig. 4. The process consists of two main components: the real-time controller for noise control and the co-processor for control filter selection. As depicted in Fig. 4, the controller operates at the sampling rate to generate the control signal, and the error signal $e(n)$ can be obtained as

$$e(n)=d(n)-x(n)*w(n)*s(n), \tag{1}$$

where $d(n)$, $x(n)$, $w(n)$ and $s(n)$ represents disturbance, reference signal, control filter and secondary path respectively, $*$ denotes the linear convolution.

Simultaneously, the reference signal and error signal at each frame $\mathbf{z}^k=[\mathbf{x}^k,\mathbf{e}^k]$ are collected and sent to the co-processor. The goal is to detect frequency changes using the reference signal and path variations using the error signal, which contains information about both primary and secondary paths. Based on the dual inputs, the CNN outputs the predicted probabilities for each frequency and temperature class as

$$(\hat{\mathbf{p}}_{frequency},\hat{\mathbf{p}}_{temperature})=CNN(\mathbf{z}^k;\Theta^*), \tag{2}$$

where $\Theta^*$ is the trained CNN parameters, $\hat{\mathbf{p}}_{frequency}=[\hat{p}_{frequency,1},\dots,\hat{p}_{frequency,F}]$ is the predicted probability for $F$ frequency classes, $\hat{\mathbf{p}}_{temperature}=[\hat{p}_{temperature,1},\dots,\hat{p}_{temperature,T}]$ is the predicted probability for $T$ temperature classes. Finally, the frequency index $i$ and the temperature index $j$ can be obtained as

$$i=\arg\max(\hat{\mathbf{p}}_{frequency}),\ j=\arg\max(\hat{\mathbf{p}}_{temperature}), \tag{3}$$

Finally, the coefficients of the selected control filter $\mathbf{w}_{ij}$ are updated at the frame rate. By facilitating cooperation between the real-time noise controller and the co-processor, the proposed method enables delayless noise control.

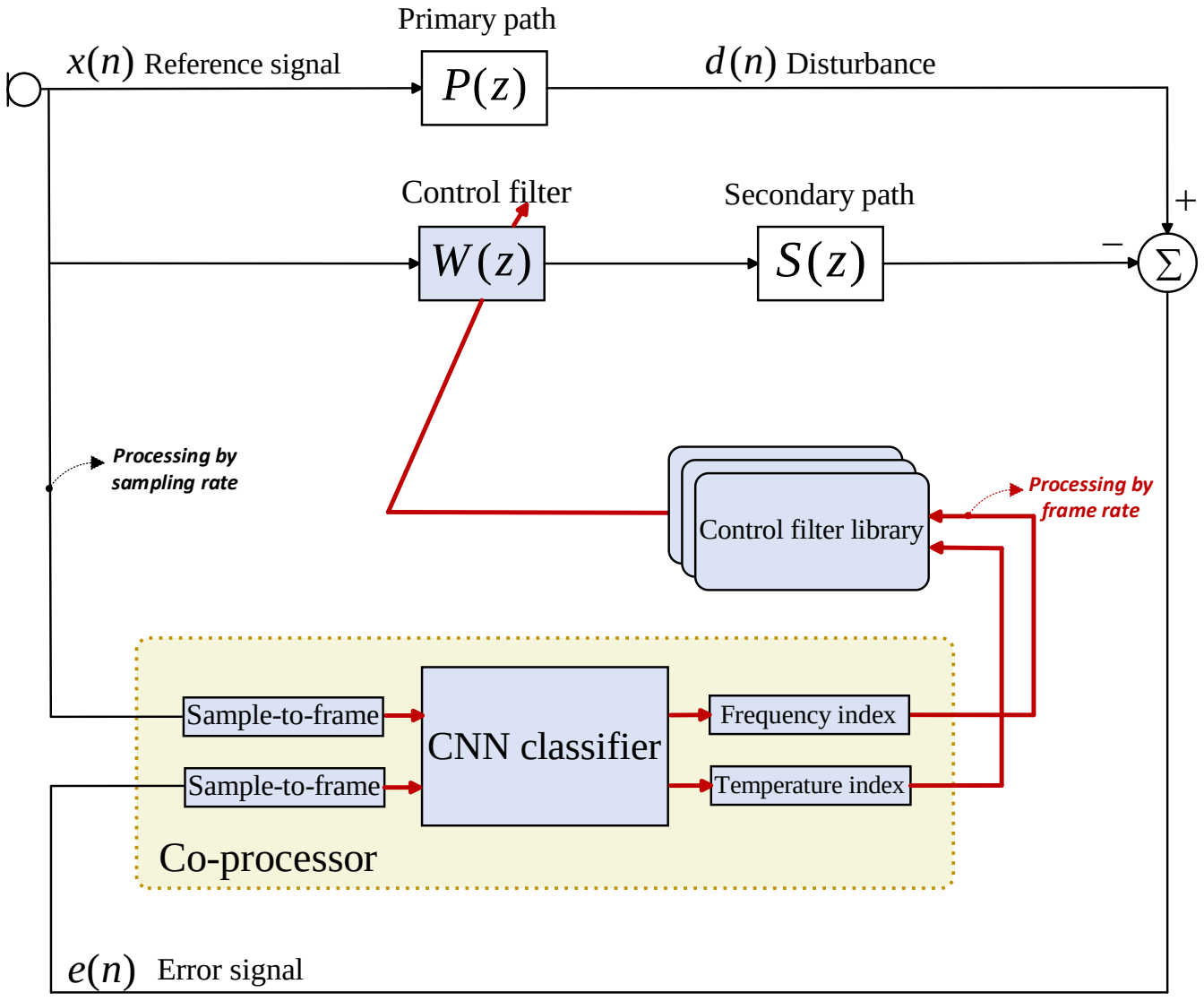

Figure 4: Block diagram of the temperature-perceptive SFANC technique.

## 3.1 Pre-trained control filter library

Before the online execution of the TP-SFANC technique, a control filter library comprising various frequency and temperature conditions must be pre-trained. Based on changes in engine shaft speed, the first component of the multi-tonal engine noise ( $f_1$ ) is assumed to vary from 108 Hz to 132 Hz. Following the method in Section 2.2, 8 multi-narrowband primary noises are synthesized across frequency ranges:

108–111 Hz, 111–114 Hz, 114–117 Hz, 117–120 Hz, 120–123 Hz, 123–126 Hz, 126–129 Hz, and 129–132 Hz. In terms of temperature, considering the varying conditions that might occur during aircraft operation, 5 temperatures (20℃, 28℃, 36℃, 44℃, 52℃) are selected, each corresponding to a unique primary and secondary path pair as described in Section 2.3. The FxLMS algorithm is then used to pre-train the control filters. As shown in Fig. 5(a), the library contains 40 control filters for online use.

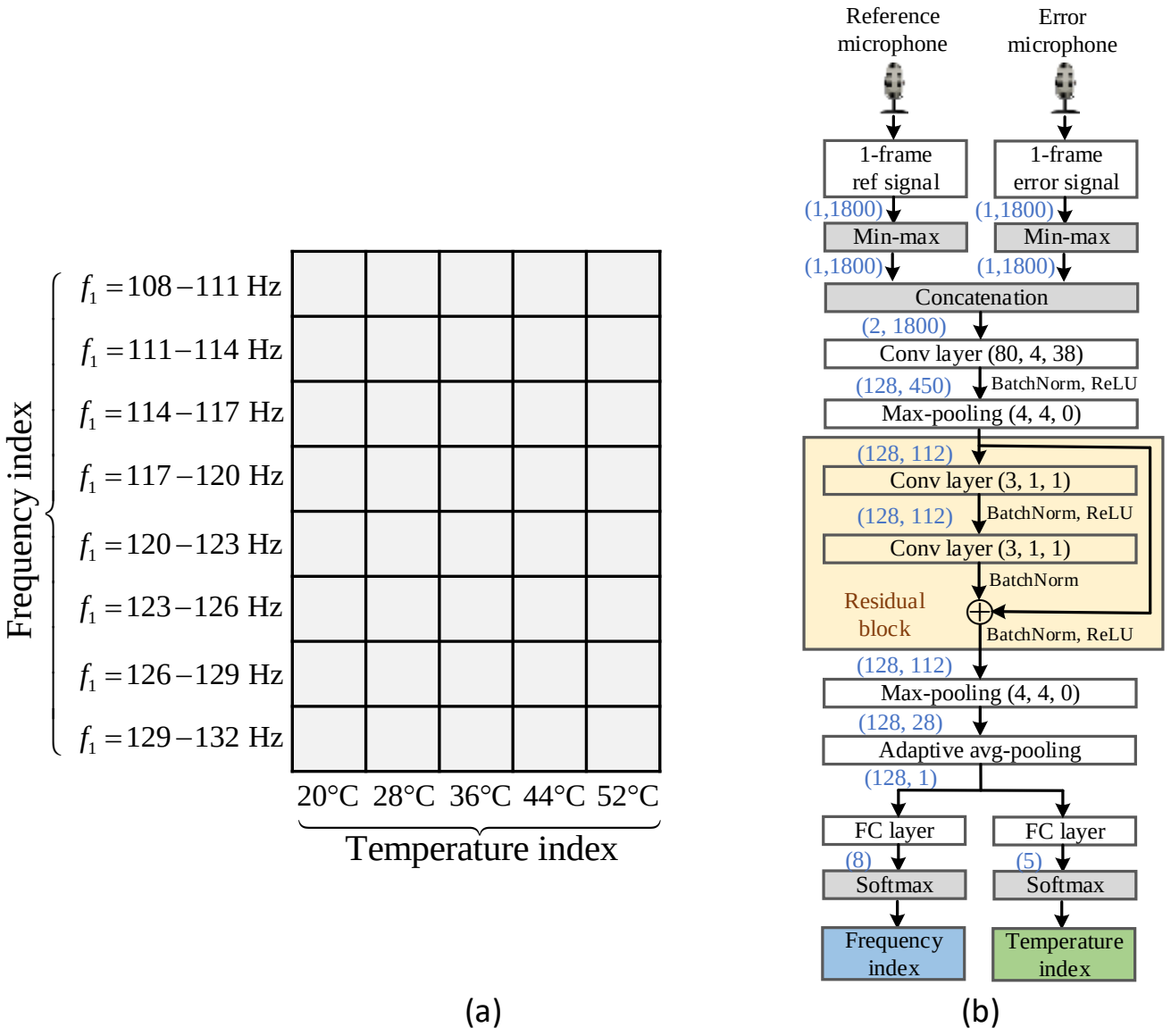

Figure 5: (a) The pre-trained control filter library and (b) block diagram of the proposed 1D CNN architecture, with convolutional layers and max-pooling layers configured as (kernel size, stride, padding).

### 3.2 1D CNN trained using a multi-task learning strategy

Fig. 5(b) shows the architecture of the proposed 1D CNN. The input consists of a one-frame reference signal and error signal, each normalized to (−1, 1) using min-max normalization and concatenated. The data pass through a convolutional layer, a max-pooling layer, and a residual block. Batch normalization and ReLU activation follow each convolution to accelerate training and maintain positive gradients. A shortcut connection is used in the residual block to simplify optimization. Another max-pooling layer precedes the adaptive average pooling to reduce computational load. The pooled features are then fed into two fully connected layers to estimate probabilities for 8 frequency and 5 temperature classes. Final predictions are made using softmax layers.

As discussed in Section 3.1, the selected control filter is determined by both the frequency index and the temperature index generated by the CNN. To achieve this, the CNN is trained using a novel multi-task learning strategy [16], enabling it to perform two classification tasks: frequency and temperature. The loss functions for both tasks are cross-entropy loss functions, denoted as $Loss_{frequency}$ and $Loss_{temperature}$, respectively. The joint loss function used to train the CNN is formulated as a weighted sum of the individual losses and is expressed as

$$Loss = \frac{Loss_{frequency}}{Loss_{frequency} + Loss_{temperature}} Loss_{frequency} + \frac{Loss_{temperature}}{Loss_{frequency} + Loss_{temperature}} Loss_{temperature}. \tag{4}$$

By utilizing this joint loss function, the CNN can dynamically adjust its focus based on the current loss of each task, thereby effectively learning from the shared representations. This multi-task learning approach is more efficient than training separate networks for each task.

# 4. Simulation results

The effectiveness of the proposed approach is evaluated on a 1×2×1 active trim panel system with synthesized multi-tonal noise, primary and secondary paths, as described in Section 2. The sampling rate is 1800 Hz, the control filter length is 1024 taps, and the CNN input frame length is 1 second.

## 4.1 Effectiveness of 1D CNN

A synthetic noise dataset is constructed for CNN training, validation, and testing. First, 1800 primary noises are generated by randomly selecting a start frequency within 108–132 Hz and a bandwidth up to 3 Hz for $f_1$, as described in Section 2.2. Primary and secondary paths corresponding to temperatures from 20℃ to 52℃ (in 0.5℃ increments) are synthesized using the state-space model in Section 2.3, with 60 path pairs randomly selected. Time-domain error signals are then computed by combining each primary noise with a path pair, using a randomly chosen control filter from the pre-trained filter library to avoid biasing the temperature classification and minimize its impact on labeling. Temperature labels are assigned by mapping each random temperature to the nearest of the 5 predefined classes (20℃, 28℃, 36℃, 44℃, 52℃). Frequency labels are determined by selecting the index of the control filter that yields the highest noise reduction among the 8 control filters in the corresponding temperature column of the pre-trained library shown in Fig. 5(a). This results in a dataset of 108000 entries, split into training, validation, and testing sets in an 8:1:1 ratio.

As shown in Table 1, the proposed 1D CNN has only 0.12 million parameters, making it suitable for deployment on embedded devices. Moreover, the proposed CNN achieves a classification accuracy of 94.48% for frequency and 95.77% for temperature, with an overall classification accuracy of 90.66% for both tasks. These results demonstrate its effectiveness in selecting appropriate control filters under varying noise and temperature conditions, enabling efficient aircraft interior noise control.

Table 1: The effectiveness of the 1D CNN used for TP-SFANC.

| Metrics | Performance |
|---|---|
| Total parameters | 0.12M |
| Classification accuracy for frequency | 94.48% |
| Classification accuracy for temperature | 95.77% |
| Classification accuracy for both frequency and temperature | 90.66% |

## 4.2 Noise cancellation of aircraft noise under time-varying frequency conditions

To evaluate TP-SFANC under time-varying frequency conditions, simulations were performed with the trim panel temperature fixed at 20℃. The $f_1$ increased from 109 Hz to 116 Hz, then to 124 Hz, with each lasting 10 seconds. The spectrograms of the time-varying frequency aircraft noise attenuated by TP-SFANC and FxLMS are shown in Fig. 6(a)–(c). The spectrograms represent the power spectral density of the noise. Additionally, the averaged noise reduction levels achieved by the two methods are presented in Fig. 6(d). As observed from Fig. 6, the TP-SFANC technique can achieve approximately 17.5 dB noise reduction within 1 second after an initial delay caused by the control filter update. In contrast, the FxLMS algorithm exhibits a gradual convergence, achieving only around 5 dB noise reduction at the 1-second mark. These results indicate that the TP-SFANC technique significantly outperforms the FxLMS algorithm in terms of both response time and noise reduction performance.

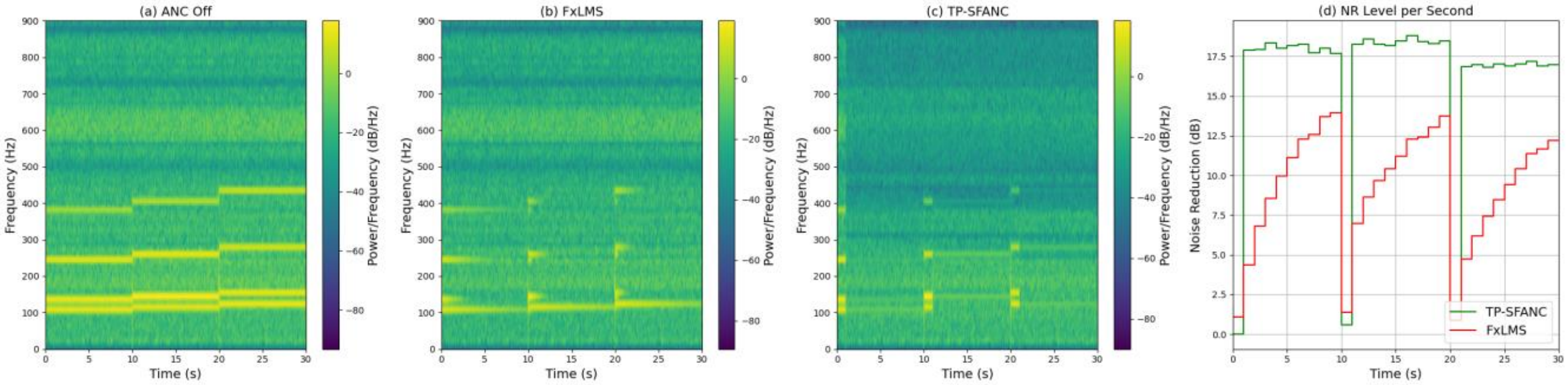

Figure 6: Spectrograms of the error signal: (a) without ANC, (b) with FxLMS, (c) with TP-SFANC, and (d) averaged noise reduction level per second, for aircraft noise under time-varying frequency.

### 4.3 Noise cancellation of aircraft noise under time-varying temperature conditions

To evaluate TP-SFANC under time-varying temperature conditions, simulations were conducted with the $f_1$ fixed at 116 Hz. The trim panel temperature increased linearly from 20℃ to 28℃ over 0–16 minutes, then fluctuated between 27.5℃ and 28.5℃ from 17–32 minutes. The magnitude of the error signal for time-varying temperatures, attenuated by FxLMS, SFANC without temperature perception, and TP-SFANC, is shown in Fig. 7(a)–(c). The SFANC without temperature perception uses a fixed filter trained at 20℃. Additionally, the averaged noise reduction levels achieved by the three methods are presented in Fig. 7(d). As illustrated in Fig. 7, the TP-SFANC technique outperforms the SFANC technique that lacks temperature perception. However, after approximately 3 minutes, the FxLMS algorithm demonstrates better performance. This can be attributed to the minimal path difference between adjacent temperatures and the gradual temperature change, allowing FxLMS to effectively track the variations. The temporary decrease in noise reduction level for TP-SFANC is due to the absence of a pre-trained control filter that precisely matches the current temperature. This limitation can be addressed by pre-training additional control filters for intermediate temperatures.

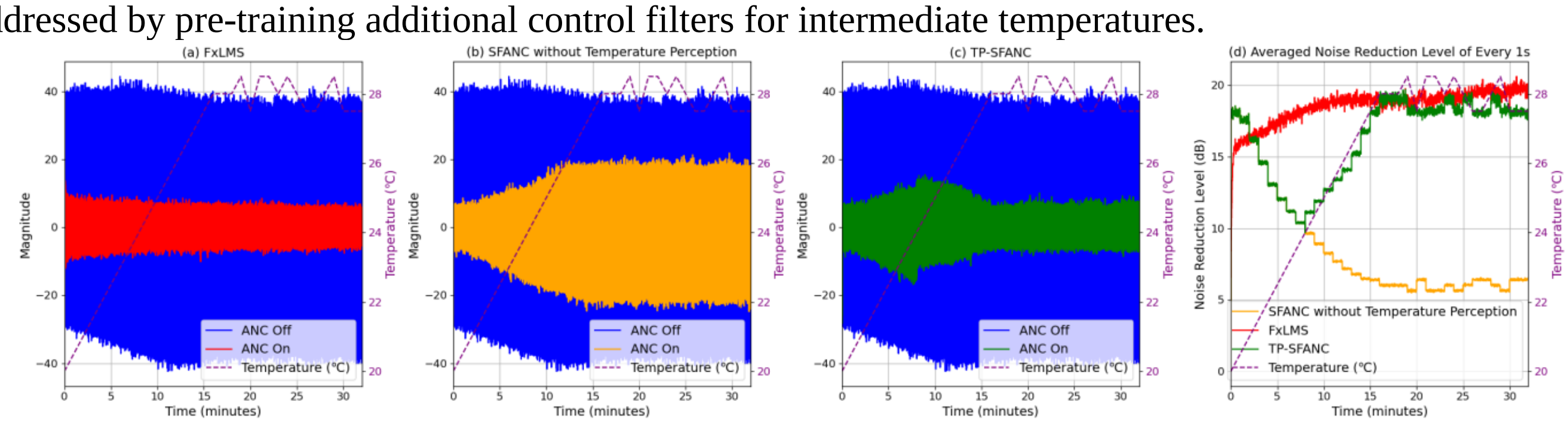

Figure 7: Magnitude of the error signal: (a) with FxLMS, (b) with SFANC without temperature perception, (c) with TP-SFANC, and (d) averaged noise reduction level per second, for active trim panel with time-varying temperature.

## 5. Conclusion

This paper presents a novel TP-SFANC technique to address multi-tonal aircraft interior noise under varying frequency and temperature conditions. The proposed method employs a lightweight 1D CNN trained using a multi-task learning strategy to dynamically select the optimal control filter based on reference and error signals. Simulation results demonstrate that TP-SFANC significantly outperforms the conventional FxLMS algorithm under varying frequency conditions and the SFANC technique under varying temperature conditions, achieving faster response times and superior noise reduction.

Despite its advantages, the effectiveness of current approach is limited to areas near the active trim panel, with reduced performance under slow temperature changes. Future work will explore a virtual sensing scheme to address these limitations.